\documentclass[12pt]{article}
\usepackage{amsmath,amssymb,cite,epsfig}
\numberwithin{equation}{section}

\begin{document}

\title{\begin{flushright}
\end{flushright}
{\bf Uniformly Accelerated Observer in Moyal Spacetime}}
\author{Nirmalendu Acharyya\footnote{nirmalendu@cts.iisc.ernet.in} \,and Sachindeo Vaidya\footnote{vaidya@cts.iisc.ernet.in}\\ 
\begin{small}Centre for High Energy Physics, Indian
Institute of Science,
Bangalore, 560012, India.
\end{small}}
\date{\empty}

\maketitle

\begin{abstract}
In Minkowski space, an accelerated reference frame may be defined as one
that is related to an inertial frame by a sequence of instantaneous Lorentz
transformations. Such an accelerated observer sees a causal horizon, and the
quantum vacuum of the inertial observer appears thermal to the accelerated
observer, also known as the Unruh effect. We argue that an accelerating frame may be similarly 
defined (i.e. as a sequence of instantaneous Lorentz transformations) in noncommutative 
Moyal spacetime, and discuss the twisted quantum field theory
appropriate for such an accelerated observer. Our analysis shows that there are
several new features in the case of noncommutative spacetime: chiral massless fields in 
$(1+1)$ dimensions have a qualitatively different behavior compared to massive fields. In addition, 
the vacuum of the inertial observer is no longer an equilibrium thermal state of the accelerating
observer, and the Bose-Einstein distribution acquires $\theta$-dependent corrections.

\end{abstract}

\section{Introduction}

Noncommutative algebras can serve as models for spacetimes at scales where quantum gravity 
effects are important, and the usual continuum description of spacetime as a manifold 
is expected to undergo significant revision. In particular, very general arguments using only 
elementary quantum uncertainties and classical gravity strongly suggest that the smooth 
spacetime structure should be replaced by a noncommutative Moyal algebra \cite{Doplicher:1994tu}. 

While the above motivation is set in a very general context, computations and specific predictions that 
involve both noncommutativity and gravity are fewer, as specific questions are harder to formulate, and 
the technical tools necessary for dealing with quantum fields on arbitrary noncommutative manifolds are 
 still in the process of being developed (see for example \cite{Bahns:2010um}). 

To this end, one would like to pose a problem in quantum field theory that is simple to formulate, 
and though does not involve 
general relativity, is perhaps the closest that one can get to discussing the nature of quantum theory in 
curved space. The Unruh effect, which is related to the quantum theory of an observer undergoing a 
uniform proper acceleration in Minkowski spacetime, is one such example. In this article, we will discuss 
the analogous question in noncommutative Groenewold-Moyal (GM) spacetime. To this end, we will 
define in a precise manner the notion of an accelerating observer and carry out detailed analysis about 
the nature of the quantum theory as formulated by this observer.

This article is organized as follows. In Section 2, we will recall  the description of an accelerated 
observer as a sequence of instantaneous Lorentz boosts. This description is particularly useful
because of the centrality of the role of Lorentz transformations, and does not rely on a 
coordinate transformation to the accelerated coordinates. In Section 3, the quantization of an 
accelerated observer in terms of modes that are boost eigenfunctions will be discussed. In section 4, 
after a brief recap of the quantum field theory in the Groenewold-Moyal (GM) plane, we shall describe its 
adaptation to that of an accelerated observer. Unusual features of massless fields will be pointed out, 
and properties related to wedge-localization and hermiticity will be discussed. In particular, we 
will describe new phenomena like loss of hermiticity of the twisted quantum field.
Section 5 consists of the main new results of our investigation: a striking consequence of the above
 is that the Minkowski vacuum is no longer a thermal state for the accelerating observer, 
but a more general mixed state. As a result, the expectation value of the number operator no longer obeys 
Bose-Einstein (B-E) distribution (as in the usual Unruh effect), and corrections to the B-E distribution 
can be explicitly computed as a perturbation series in $\theta$. Finally, we will study the above 
result with various choices of $\theta^{\mu\nu}$ and show that the correction to the B-E distribution
is non-zero unless all components of $\theta^{\mu\nu}$ are zero. In section 6, we will conclude with 
a discussion of our results. The appendices contain a discussion of eigenfunctions of the boost 
operator in various dimensions, for massless as well as massive cases.

\section{Kinematics of a Uniformly Accelerating Observer  \label{u_a_o}}

Let us remind ourselves of the description of the accelerated observer in commutative 
Minkowski spacetime ${\cal M}^{1,3}$. This description \cite{hughes} takes full advantage of the 
action of the Lorentz group, and avoids transformation to coordinates of the accelerated observer.
In $(d+1)$ dimensional Minkowski spacetime $\mathcal{M}^{1,d}$, events are labeled by coordinates 
$(x^{0},\vec{x})$ or $(t,\vec{x})$ where $\vec{x}$ is a $d$-dimensional spatial vector.

Let $\mathcal{L}_{+}^{\uparrow}$ be the group of proper Poincar\'e transformation of 
${\cal M}^{1,3}$, and $T$ and $\Lambda_{\hat{n}}$ the one-parameter subgroups corresponding to 
time translations and Lorentz boosts along a direction $\hat{n}$ respectively.  Consider a point observer 
uniformly accelerating (with respect to an  inertial frame) along the spatial direction $\hat{n}$ with 
proper four-acceleration $a^{\mu}$, satisfying
\begin{equation}
a_{\mu}a^{\mu}= -a^{2}, \quad a =\rm{constant}.
\end{equation}
If $\tau$ is the proper time of this accelerated observer, we shall choose the initial condition that at 
$t=\tau=0$, the observer is at rest. The world line of such an observer is then given by
\begin{eqnarray}
t &=& \rho \sinh(a\tau), \quad x_{\|}=\rho \cosh(a\tau), \quad
\vec{x}_{\perp}= \mathrm{constant}, \quad {\rm where} \label{rind} \\ 
\vec{x} &=& (\vec{x}\cdot\hat{n})\hat{n} +\vec{x}_{\perp} = x_{\|}
\hat{n} + \vec{x}_{\perp}.
\label{gen_vec}
\end{eqnarray}
These world lines are hyperbolas in $x_{\|}-t$ plane given by $
x_{\|}^{2}-t^{2}=\rho^{2}$ and from (\ref{rind}) it is easy to see
that $a_{\mu}a^{\mu}= -\rho^{2}$, implying that $\rho=a^{-1}$.

For any $a>0$, the world lines (\ref{rind}) of the observer  lie in the
open sub-manifold $R$ of ${\cal M}^{1,3}$ while for any $a<0$, the world
lines lie in the open sub-manifold $L $ of ${\cal M}^{1,3}$. The
sub-manifolds $R$ and $L$ are defined as
\begin{equation}
R=\{t, x_{\|}: x_{\|}>|t|\}, \quad L=\{t, x_{\|}: x_{\|}< -|t|\} .
\end{equation}
The motion given by (\ref{rind}) is generated by a Lorentz boost of magnitude $a\tau$
of the event $(0,a^{-1},\vec{0})$ about
$(0,0,\vec{x}_{\perp})$\footnote{\small events are specified as
  $(t,x_{\|},\vec{x}_{\perp})$}. In other words, the motion of the
accelerated observer is generated by $a \vec{\mathcal{K}}\cdot
\hat{n} \equiv a \mathcal{K}$, where $\mathcal{K}^{i} $'s are the
generators of Lorentz boosts.

It is easy to construct the Rindler coordinates $(\tau,\xi,\vec{x}_{\perp})$
appropriate for region $R$ (the right Rindler wedge). The
transformation equations between the Rindler coordinates and Minkowski
coordinates in $R$ are
\begin{equation}
t=a^{-1}e^{a\xi}\sinh(a\tau), \quad x_{\|}=a^{-1}e^{a\xi}\cosh(a\tau),
\quad \vec{x}_{\perp}=\vec{x}_{\perp}.
\label{transformation_r}
\end{equation}

Similarly, for the left Rindler wedge, the coordinates $(\bar{\tau},\bar{\xi},
\vec{x}_{\perp})$ are related to $(t,x_{\|},\vec{x}_{\perp})$ as
\begin{eqnarray}
t=-a^{-1}e^{a\bar{\xi}}\sinh(a\bar{\tau}), \quad
x_{\|}=a^{-1}e^{a\bar{\xi}}\cosh(a\bar{\tau}), \quad
\vec{x}_{\perp}=\vec{x}_{\perp}.
\end{eqnarray}

The one-parameter subgroups  $T$ and  $\Lambda_{\hat{n}}$ act on the
coordinates $(t,x_{\|},\vec{x}_{\perp})$ as follows:
\begin{eqnarray}\left.
\begin{array}{l l l} 
T(\beta)(t,x_{\|},\vec{x}_{\perp}) &=& (t+\beta,x_{\|},\vec{x}_{\perp}), \\
\Lambda_{\hat{n}}(a\alpha) (t,x_{\|},\vec{x}_{\perp})&=&(t \cosh(a\alpha)+x_{\|}
\sinh(a\alpha),t \sinh(a\alpha) + x_{\|}\cosh(a\alpha),\vec{x}_{\perp}).
\end{array}\right.
\label{nboost}
\end{eqnarray}
Since both $R$ and $L$ are preserved under $\Lambda_{\hat{n}}$, it is meaningful
to define $\Lambda^{R}_{\hat{n}}$ and $\Lambda^{L}_{\hat{n}}$, the restrictions of
$\Lambda_{\hat{n}}$ to $R$ and $L$ respectively.  Using (\ref{rind}) and
(\ref{nboost}), we see that
\begin{equation}
\Lambda^{R}_{\hat{n}}(a\alpha)(\rho,a\tau,\vec{x}_{\perp})=(\rho,a(\tau+\alpha),\vec{x}_{\perp}),
\quad {\rm for} \quad (\rho,\tau,\vec{x}_{\perp}) \in R.
\label{timetranslator}
\end{equation}
Thus $\Lambda^{R}_{\hat{n}}$ corresponds to time-translations for the uniformly
accelerating observer in $R$.  A similar remark holds for
$\Lambda^{L}_{\hat{n}}$ as well.  This feature is special to the Rindler
observer, and is used to construct the quantum field for the uniformly
accelerating observer, as we shall see in the next section.

\section{QFT in an Accelerated Frame for $\theta^{\mu\nu}=0$}

Let us first recall the description of  quantum field theory in an inertial frame. 
 The quantum field $\Phi_{0}(x^{0},\vec{x}) $ on $\mathcal{M}^{1,d}$ is an operator-valued
distribution, which, when expanded in terms of plane-waves in $\mathcal{M}^{1,d}$, 
can be written as
\begin{equation}
\Phi_{0}(x^{0},\vec{x})=\int\frac{d^{d}\vec{k}}{(2\pi)^{d}(2k^{0})}
\left(b_{\vec{k}} f_{\vec{k}}(x^{0},\vec{x}) +b_{\vec{k}}^{\dagger}
f_{\vec{k}}^{*}(x^{0},\vec{x})\right)
\label{field_expansion}
\end{equation}
where $f_{\vec{k}}(t,\vec{x})= e^{-i(k^{0}t-\vec{k}\cdot\vec{x})}$ and
$k^{\mu} \equiv (k^{0},\vec{k})$ with $k_{\mu}k^{\mu}=m^{2}$.

The annihilation and creation operators
$b_{\vec{k}},b_{\vec{k}}^{\dagger} $ satisfy
 \begin{eqnarray}
[b_{\vec{k}},b_{\vec{k}^{\prime}}^{\dagger}]&=&(2\pi)^{d}(2k_{0})
\delta^{d}(\vec{k}-\vec{k}^{\prime})
\label{commutator10}
 \end{eqnarray}
with all other commutators vanishing. The operators
$b_{\vec{k}},b_{\vec{k}}^{\dagger} $ act on the  Fock space,
which has a distinguished state $|0_{M}\rangle $ (known as the vacuum
state), that is annihilated by $b_{\vec{k}}$:
\begin{equation}
 b_{\vec{k}}|0_{M}\rangle=0 \quad \forall  \,\, \vec{k}.
 \end{equation}
The massless scalar field in $\mathcal{M}^{1,1}$ has a
special feature that we will remind ourselves of, as it will play an
important role in our subsequent discussion. (Strictly speaking, the
massless field in $\mathcal{M}^{1,1}$ is problematic because the
two-point correlation function has an infrared divergence, but this
subtlety is not relevant to our discussion). In terms of its modes,
the massless quantum field can be expanded as
\begin{eqnarray}
\Phi_{0}(x^{0},x^{1})=\int_{-\infty}^{\infty}\frac{dk^{1}}{(2\pi)(2k^{0})}[b_{k^{1}}
  f_{k^{1}}(x^{0},x^{1}) +b_{k^{1}}^{\dagger}
  f_{k^{1}}^{*}(x^{0},x^{1}) ], \quad {\rm with} \quad k^{0}=|k^{1}|.
\label{field_expansion1}
\end{eqnarray}
The field $\Phi_{0}(x^{0},x^{1})$ is the sum of two
linearly independent non-interacting chiral fields -- a right-mover
$\Phi_{0}^{r}$ and a left-mover $\Phi_{0}^{l}$. In terms of the
{\it lightcone coordinates} $U \equiv x^{0}-x^{1}$ and $V \equiv x^{0}+ x^{1}$,
 the quantum field expansion is
\begin{eqnarray}
\Phi_{0}(x^{0},x^{1})&=&\int_{0}^{\infty}\frac{dk^{1}}{(2\pi)(2k^{1})}
\left( b_{k^{1}} e^{-ik^{1}U} +b_{k^{1}}^{\dagger}e^{ik^{1}U} +
 b_{-k^{1}} e^{-ik^{1}V} +b_{-k^{1}}^{\dagger}e^{ik^{1}V}
\right)\\ &\equiv&\Phi_{0}^{r}(U)+\Phi_{0}^{l}(V)
\label{field_expansion2}
\end{eqnarray}
where we have used the relation $k^{0}=|k^{1}|$. Since the left- and
right-movers do not interact (at the level of the free field
expansion), it is possible to discuss physical situations involving a pure
left-mover (or a pure right-mover) only, by simply dropping the quantum
field of the other chirality.

In the accelerated frame, the operators $b^{\dagger}_{\vec{k}}$ do not 
create energy eigenstates. Rather, the natural choice for expanding the
quantum field is the basis of the eigenfunctions of the generator of
Rindler-time translations. 

In the previous section, we argued that the Lorentz boost operator
${\mathcal{K}}(\equiv \vec{\mathcal{K}}\cdot \hat{n})$ restricted to 
one of the Rindler wedges is the
generator of time translations in the accelerated frame. We will
denote this restricted operator as $\mathcal{K}^{R}$ (or
$\mathcal{K}^{L}$). The eigenfunctions
$\{\varphi_{\omega,\vec{k}_\perp,R (L)}(t,x_{\|},\vec{x}_{\perp})\}$
of $\mathcal{K}^{R(L)}$ (see Appendix A) form a complete orthonormal
basis and thus can be used as the modes of the quantum field 
in the accelerated frame.

Since $\varphi_{\omega,\vec{k}_{\perp},R}$ vanishes in $L$ (and
 $\varphi_{\omega,\vec{k}_{\perp},L}$ vanishes in $R$), we can write 
$\Phi_{0}(t,x_{\|},\vec{x}_\perp)
= \Phi_{0}^{L}(t,x_{\|},\vec{x}_\perp) +
\Phi_{0}^{R}(t,x_{\|},\vec{x}_\perp)$ where
\begin{eqnarray}
\Phi_{0}^{R}(t,x_{\|},\vec{x}_\perp) = \int d^{d-1}\vec{k}_\perp\int_{0}^{\infty} 
\frac{d\omega}{(2\pi)^{d}(2\omega)}\left( a_{\omega,\vec{k}_\perp }^{R} 
\varphi_{\omega,\vec{k}_\perp,R}(t,x_{\|},\vec{x}_\perp) + 
a^{R\dagger}_{\omega,\vec{k}_\perp}\varphi^{\ast}_{\omega,\vec{k}_\perp,R}
(t,x_{\|},\vec{x}_\perp) \right),\\
\Phi_{0}^{L}(t,x_{\|},\vec{x}_\perp) = \int d^{d-1}\vec{k}_\perp\int_{0}^{\infty} 
\frac{d\omega}{(2\pi)^{d}(2\omega)}\left( a_{\omega,\vec{k}_\perp}^{L} 
\varphi^{\ast}_{\omega,\vec{k}_\perp,L}(t,x_{\|},\vec{x}_\perp) + 
a^{L\dagger}_{\omega,\vec{k}_\perp}\varphi_{\omega,\vec{k}_\perp,L}
(t,x_{\|},\vec{x}_\perp)\right).
\end{eqnarray}
Here $\varphi_{\omega,\vec{k}_\perp,R}(t,x_{\|},\vec{x}_\perp)$ and
$\varphi_{\omega,\vec{k}_\perp,L}(t,x_{\|},\vec{x}_\perp)$ are given
by (\ref{eigenkrd}) and (\ref{eigenkld}) respectively, and the
operators $a_{\omega,\vec{k}_\perp }^{R(L)},a^{R(L)
  \dagger}_{\omega,\vec{k}_\perp}$ satisfy
\begin{eqnarray}
 [a_{\omega,\vec{k}_\perp}^{R(L)},a_{\omega^{\prime},\vec{k}_\perp^{\prime}}^{R(L) \dagger}]=(2\pi)^{d}(2\omega)\delta(\omega-\omega^{\prime})\delta^{d-1}(\vec{k}_{\perp}-\vec{k}_{\perp}^{\prime}),
\label{rind-commutator5}
\end{eqnarray}
with all other  commutators vanishing.

Even in the Rindler wedges $R$ and $L$, the massless quantum field in
$(1+1)$ dimensions can be split into left and right-movers
$\Phi_{0}^{R(L), \lambda}(t,x_{\|})$. We use the label $\lambda$ to distinguish 
these two cases, with $\lambda=1$ for the right-mover and  $\lambda=2$ for the 
left-mover.
For the massless quantum field $\Phi_{0}(t,x_{\|})$ in
$\mathcal{M}^{1,1}$, we have $\Phi_{0}(t,x_{\|}) =
\Phi_{0}^{L}(t,x_{\|})+\Phi_{0}^{R}(t,x_{\|})$ where
\begin{eqnarray} 
\Phi_{0}^{R}(t,x_{\|})&=& \sum_{\lambda=1}^{2}\int_{0}^{\infty} 
\frac{d\omega}{(2\pi)(2\omega)} \left( a_{\omega,\lambda}^{R} 
\varphi_{\omega,\lambda,R}(t,x_{\|}) + 
a^{R\dagger}_{\omega,\lambda}\varphi^{\ast}_{\omega,\lambda,R}(t,x_{\|}) \right)
\label{massless_field_RRW}, \\
\Phi_{0}^{L}(t,x_{\|})&=& \sum_{\lambda=1}^{2}\int_{0}^{\infty} 
\frac{d\omega}{(2\pi)(2\omega)} \left( a_{\omega,\lambda}^{L} 
\varphi^{\ast}_{\omega,\lambda,L}(t,x_{\|}) + 
a^{L\dagger}_{\omega,\lambda}\varphi_{\omega,\lambda,L}(t,x_{\|}) \right).
\label{massless_field_LRW}
\end{eqnarray}
 
The $a_{\omega,\lambda}^{R},a^{R\dagger}_{\omega,\lambda}$ satisfy the
commutation relations
\begin{eqnarray}
[a_{\omega,\lambda}^{R},a^{R\dagger}_{\omega^{\prime},\lambda^{\prime}}] = (2\pi)(2\omega)\delta_{\lambda\lambda^{\prime}}\delta(\omega-\omega^{\prime})= [a_{\omega,\lambda}^{L},a^{L\dagger}_{\omega^{\prime},\lambda^{\prime}}],
\end{eqnarray}
with all other commutators vanishes.


The discussion of a massive quantum field in $(1+1)$ dimensional Rindler spacetime is very similar to above: the 
only difference is that the eigenfunctions of the boost operator are different and there is 
no label $\lambda$. The field in the right Rindler wedge has the expansion
\begin{eqnarray}
\Phi_{0}^{R}(t,x_{\|})&=&\int_{0}^{\infty}\frac{d\omega}{(2\pi)(2\omega)}\left(a_{\omega}^{R} \varphi_{\omega,R}(t,x_{\|})+a_{\omega}^{R \dagger} \varphi_{\omega,R}^{*}(t,x_{\|})\right) 
\label{rind-minkcoord}
\end{eqnarray}
with the operators $a_{\omega}^{R}$ and $a_{\omega}^{R \dagger}$ satisfying the 
relation (\ref{rind-commutator5}).
The explicit form of $\varphi_{\omega,R}$ is given in (\ref{eigenfunc}).

The generalization to $(3+1)$ dimensions is straightforward. The quantum field is
\begin{eqnarray}
\Phi_{0}^{R}(t,x_{\|},\vec{x}_\perp)=\int\frac{d^{2}\vec{k}_{\perp}d\omega}{(2\pi)^3 (2\omega)}\left(a_{\omega,\vec{k}_\perp}^{R} \varphi_{\omega,\vec{k}_\perp,R}(t,x_{\|},\vec{x}_\perp)+a_{\omega,\vec{k}_\perp}^{R \dagger} \varphi_{\omega,\vec{k}_\perp,R}^{ *}(t,x_{\|},\vec{x}_\perp)\right)
\label{rind-minkcoord2}
\end{eqnarray}
where $ \varphi_{\omega,\vec{k}_\perp,R}$ is given by (\ref{eigenfunc2}).
The operators $a_{\omega,\vec{k}_\perp}^{R}$ and $a_{\omega,\vec{k}_\perp}^
{R \dagger}$ satisfy the cannonical commutation relation (\ref{rind-commutator5}).

The vacuum $|0_{R}\rangle$ of the Rindler observer (in, say, $R$) is annihilated by 
$a_{\omega,\vec{k}_\perp}^{R}$:
\begin{equation}
 a_{\omega,\vec{k}_\perp}^{R}|0_{R}\rangle=0
\end{equation}
and  $\hat{\mathcal{N}}_{\omega,\vec{k}_\perp,R}=a_{\omega,\vec{k}_\perp}^{R \dagger}a_{\omega,\vec{k}_\perp}^{R}$ 
is the number operator which counts the number of Rindler particles with eigenvalue $\omega$ and 
label $\vec{k}_{\perp}$.
 The operator $b_{\vec{k}}^{\dagger}$ and $a_{\omega,\vec{k}_{\perp}}^{R \dagger}$  are 
related by a Bogoliubov transformation (see for example \cite{crispino})
\begin{eqnarray}
 a_{\omega,\vec{k}_\perp}^{R}=\int_{-\infty}^{\infty} \frac{dk_{\|}}{(2\pi)(2k_{0})}\left(\alpha_{\omega,\vec{k}_\perp}^{ R}(k_{\|})b_{\vec{k}}+\beta_{\omega,\vec{k}_\perp}^{ R *}(k_{\|})b_{\vec{k}}^{\dagger}\right)
\label{bogolibov2}
\end{eqnarray}
where  $\alpha_{\omega,\vec{k}_\perp}^{ R}(k_{\|}),\beta_{\omega,\vec{k}_\perp}^{ R *}(k_{\|}) \in {\mathbb C}$  
are the Bogoliubov coefficients which can be evaluated explicitly.

In $(1+1)$ dimensional massless case, $\hat{\mathcal{N}}_{\omega,\lambda,R} = a_{\omega,\lambda}^{R \dagger}a_{\omega,\lambda}^{R}$ counts the number of 
particles with eigenvalue $\omega$ and label $\lambda$.
The corresponding bogoliubov transformations are
\begin{eqnarray}
\left.
\begin{array}{l l}
a_{\omega,1}^{R}=\int_{0}^{\infty} \frac{dk_{\|}}{(2\pi)(2k_{0})}\left(\alpha_{\omega,1}^{ R}(k_{\|})b_{k_{\|}}+\beta_{\omega,1}^{ R *}(k_{\|})b_{k_{\|}}^{\dagger}\right) \\
a_{\omega,2}^{R}=\int_{0}^{\infty} \frac{dk_{\|}}{(2\pi)(2k_{0})}\left(\alpha_{\omega,2}^{ R}(k_{\|})b_{-k_{\|}}+\beta_{\omega,2}^{ R *}(k_{\|})b_{-k_{\|}}^{\dagger}\right) \\
\end{array}
\right.
\label{bogolibov1}
\end{eqnarray}
where  $\alpha_{\omega,\lambda}^{ R}(k_{1}),\beta_{\omega,\lambda}^{ R *}(k_{1}) \in {\mathbb C}$. Again the Bogoliubov coefficients can be evaluated explicitly.
 
The Minkowski vacuum state $|0_{M}\rangle$ is not the vacuum of either of the Rindler 
wedges. Owing to equations (\ref{bogolibov2}) and (\ref{bogolibov1}) it can be shown that $\langle0_{M}| \hat{\mathcal{N}}_{\omega,\vec{k}_\perp,R}|0_{M}
\rangle$ is the usual Bose-Einstein distribution.
 

\section{QFT of an Accelerated Observer for $\theta^{\mu\nu} \neq 0$}

The GM plane is a specific noncommutative deformation of the ordinary 
Minkowski spacetime in which the spacetime coordinate functions $x^{\mu}$ become 
(self-adjoint) operators $\hat{x}^{\mu}$ satisfying the commutation relations
\begin{eqnarray}
 [\hat{x}^{\mu},\hat{x}^{\nu}]=i\theta^{\mu\nu}, \quad \mu,\nu = 0,1,\cdots, d.
\label{coord_commutator}
\end{eqnarray}
The matrix $\theta^{\mu\nu}$ is a real antisymmetric constant matrix of dimension of length-squared 
$[L]^2$. It is the scale of the smallest area in the $x^{\mu}-x^{\nu}$ plane below which no length 
scale can be probed, and the spacetime becomes ``fuzzy'' at this scale.

The algebra of functions ${\cal A}_\theta({\mathbb R}^{d+1})$ on the GM
plane consists of smooth functions on ${\mathbb R}^{d+1}$, with the
multiplication map
\begin{eqnarray}
m_\theta: {\cal A}_\theta ({\mathbb R}^{d+1}) \otimes {\cal A}_\theta
({\mathbb R}^{d+1}) &\rightarrow& {\cal A}_\theta ({\mathbb R}^{d+1})\,,
\nonumber \\
\alpha \otimes \beta &\rightarrow& \alpha \;e^{\frac{i}{2}
  \overleftarrow{\partial}_\mu \theta^{\mu \nu}
  \overrightarrow{\partial}_\nu} \ \beta := \alpha \ast \beta \,.
\label{starmult}
\end{eqnarray}

Equivalently,
\begin{equation}
m_\theta (\alpha \otimes \beta) = m_0 [F_\theta \alpha \otimes
\beta] 
\label{starmult1}
\end{equation}
where $m_0$ is the point-wise multiplication map, and $F_\theta = e^{\frac{i}{2} \partial_\mu \otimes \theta^{\mu \nu} \partial_\nu}$ the twist element.

The usual action of the Lorentz group $\mathcal {L}_{+}^{\uparrow}$ is not compatible
with $\ast$-multiplication: transforming $\alpha$ and $\beta$
separately by an arbitrary group element $\Lambda \in \mathcal{ L}_{+}^{\uparrow}$ and
then $\ast$-multiplying them is not the same as transforming their
$\ast$-product. However, with a new twisted coproduct $\Delta_\theta(\Lambda)$, the action of the Lorentz group is now compatible \cite{chaichian}. 

In quantum theory, we require that (anti)-symmetrization of states describing identical particles 
be compatible with the symmetries of the underlying spacetime. The twisted coproduct 
$\Delta_\theta(\Lambda)$ forces us to alter the notion of (anti)-symmetrization as well \cite{bmpv,bal2}.

For a quantum field $\Phi_{\theta}(x^{0},\vec{x})$ on GM spacetime
\begin{eqnarray}
\Phi_{\theta}(x^{0},\vec{x}) = \int\frac{d^{d}\vec{k}}{(2\pi)^{d}(2k^{0})} \left(b_{\vec{k}}^{\theta} 
f_{\vec{k}}(x^{0},\vec{x}) + b_{\vec{k}}^{\theta\dagger} f_{\vec{k}}^{*}(x^{0},\vec{x}) \right), 
\quad k^{\mu}k_{\mu}=m^{2}
\label{field_expansion_twisted}
\end{eqnarray}
this implies a specific deformation of the algebra of creation/annihilation operators:
\begin{eqnarray} 
b^{\theta\dagger}_{\vec{k}} b^{\theta\dagger}_{\vec{k}^{\prime}} &=& e^{ik_{\mu}\theta^{\mu\nu} 
k_{\mu}^{\prime}} b^{\theta\dagger}_{\vec{k}^{\prime}} b^{\theta\dagger}_{\vec{k}}, \\
b^{\theta}_{\vec{k}} b^{\theta}_{\vec{k}^{\prime}} &=& e^{ik_{\mu}\theta^{\mu\nu}k_{\mu}^{\prime}} 
b^{\theta}_{\vec{k}^{\prime}} b^{\theta}_{\vec{k}} \\
b^{\theta}_{\vec{k}} b^{\theta\dagger}_{\vec{k}^{\prime}} &=& 
e^{-ik_{\mu}\theta^{\mu\nu}k_{\nu}^{\prime}}b^{\theta\dagger}_{\vec{k}^{\prime}}
b^{\theta}_{\vec{k}}+(2\pi)^d 2k^{0} \delta^{d} (\vec{k}-\vec{k}')
\end{eqnarray}

Twisted and the untwisted creation-annihilation operator are related to each other by a dressing 
transformation (Grosse-Faddeev-Zamolodchikov algebra) \cite{Grosse,zamalodchikov,faddeev}
\begin{equation}
  b^{\theta}_{\vec{k}}= b_{\vec{k}}e^{-\frac{i}{2}k_{\mu}\theta{\mu\nu}\mathcal{P}_{\nu}}
\label{dress1}
\end{equation}
where $\mathcal{P}_{\nu}$ is the total four momentum operator given by
\begin{equation}
 \mathcal{P}_{\mu}=\int \frac{d^{d}\vec{k}}{(2\pi)^{d}(2k^{0})^{d}}(b^{\dagger}_{\vec{k}}b_{\vec{k}})k_{\mu} 
 = \int \frac{d^{d}\vec{k}}{(2\pi)^{d}(2k^{0})^{d}}(b^{\theta\dagger}_{\vec{k}}b^{\theta}_{\vec{k}}) k_{\mu}
\label{field_mom}.
\end{equation}
The map (\ref{dress1}) is invertible.
\newline The twisted statistics can also be accounted for by writing 
\begin{eqnarray}
 \Phi_{\theta}(x)=\Phi_{0}(x)e^{\frac{1}{2}\overleftarrow{\partial}_{\mu}\theta^{\mu\nu}\mathcal{P}_{\nu}} 
\equiv \Phi_{0}(x)e^{\frac{1}{2}\overleftarrow{\partial} \wedge \mathcal{P}}
\label{dress2}
\end{eqnarray}
where $\overleftarrow{\partial} \wedge \mathcal{P}=\overleftarrow{\partial}_{\mu}
\theta^{\mu\nu}\mathcal{P}_{\nu}$.
The discussion is valid for all dimensions $d>0$ and for both massive and massless case.

Let us consider a situation in which there is  only a right mover:
\begin{eqnarray}
 \Phi_{\theta}^{r}(U)=\Phi_{0}^{r}(U) e^{\frac{1}{2}\overleftarrow{\partial}_{\mu} 
 \theta^{\mu\nu}\mathcal{P}_{\nu}}.
\label{twist_single_chiral1}
\end{eqnarray}
 From (\ref{field_expansion2}) and (\ref{twist_single_chiral1}) we get 
 \begin{equation}
 \Phi_{\theta}^{r}(U) = \int_{0}^{\infty}\frac{dk^{1}}{(2\pi)(2k^{0})} \left( b_{k^{1}}e^{-\frac{i}{2}\theta 
 k^{1}(\mathcal{P}_{0}+\mathcal{P}_{1})} e^{-ik^{1}U} + b^{\dagger}_{k^{1}}
 e^{\frac{i}{2}\theta k^{1}(\mathcal{P}_{0}+\mathcal{P}_{1})} e^{ik^{1}U} \right),
\end{equation}
where
\begin{eqnarray}
 \mathcal{P}_{\mu}&=&\int_{0}^{\infty} \frac{dk^{1}}{(2\pi)(2k^{0})}(b^{\dagger}_{k^1}b_{k^1})k_{\mu}.
\label{field_mom_massless}
\end{eqnarray}
Since $k_{0}=-k_{1}$ for a right-mover, we find that $(\mathcal{P}_{0} + \mathcal{P}_{1})=0 $, and 
hence the twisted chiral field  $\Phi_{\theta}^{r}(U)$ is same as the untwisted chiral field  
$\Phi_{0}^{r}(U)$.

An identical argument shows that $\Phi_{\theta}^{l}(U)$ is same as the untwisted chiral field  $\Phi_{0}^{l}(U)$.

On the other hand, if we need both chiralities to be present, then the twisted field is indeed distinct 
from its untwisted counterpart. This is easy to demonstrate using the total momentum operator
 \begin{eqnarray}\left.
\begin{array} {l l l l}
  \mathcal{P}_{0}&=&\int_{0}^{\infty} \frac{dk^{1}}{(2\pi)(2k^{0})}(b^{\dagger}_{k^1}b_{k^1}+b^{\dagger}_{-k^1}b_{-k^1})k_{0}, \\
\mathcal{P}_{1}&=&\int_{0}^{\infty} \frac{dk^{1}}{(2\pi)(2k^{0})}(b^{\dagger}_{k^1}b_{k^1}-b^{\dagger}_{-k^1}b_{-k^1})k_{1}. 
\end{array}\right.
\label{field_mom1}
 \end{eqnarray}

As we saw in Section \ref{u_a_o}, the motion of a uniformly accelerated observer in Minkowski 
spacetime is generated by Lorentz boost $\mathcal{K} \equiv\vec{\mathcal{K}}\cdot \hat{n}$.
The quantum field theory in the accelerated frame can be formulated using this observation,
directly in terms of Minkowski coordinates $(t,x_{\|},\vec{x}_{\perp})$. This circumvents the need 
to change the coordinates to that of the accelerated observer (\ref{rind}), and finding modes of the 
Klein-Gordon equation in the new coordinates, and subsequent quantization.
As the Poincar\'e group is an automorphism of the GM plane, we shall simply define an accelerated frame 
in GM plane as one related to an inertial frame via a sequence of instantaneous Lorentz 
transformation generated by $\mathcal{K}$.

The immediate and natural question, given that an accelerated frame can now be defined in the 
GM plane, is the nature of this noncommutative quantum field theory. Our strategy is to use the twist map   
 (\ref{dress2}) to define this quantum field as follows:
\begin{eqnarray}\left.
\begin{array}{l l l}
\Phi_{\theta}^{R}(t,\vec{x})=\Phi_{0}^{R}(t,\vec{x})e^{\frac{1}{2}\overleftarrow{\partial}\wedge\mathcal{P}} \\
\Phi_{\theta}^{L}(t,\vec{x})=\Phi_{0}^{L}(t,\vec{x})e^{\frac{1}{2}\overleftarrow{\partial}\wedge\mathcal{P}} \\
\end{array}\right\}.
\label{twist_rindler}
\end{eqnarray}
To construct the field $\Phi_{\theta}^{R}$, we simply need to start with $\Phi_{0}^{R}$ and twist it as above.

For the massless case, (\ref{massless_field_RRW}) and (\ref{twist_rindler}) give
\begin{eqnarray}
\Phi_{\theta}^{R}(t,x_{\|})&= [\Phi_{0}^{R,1}(t,x_{\|})+\Phi_{0}^{R,2}(t,x_{\|})]
e^{\frac{1}{2}\overleftarrow{\partial}\wedge\mathcal{P}}&\equiv\Phi_{\theta}^{R,1}(t,x_{\|}) + 
\Phi_{\theta}^{R,2}(t,x_{\|})
\end{eqnarray}
where $\Phi_{0}^{R,1}(t,x_{\|}),\Phi_{0}^{R,2}(t,x_{\|})$ are the right and the left movers in 
commutative right Rindler wedge. 
$\Phi_{\theta}^{R,1}(t,x_{\|})$ is the 
right moving part of the twisted quantum field $\Phi_{\theta}^{R}(t,x_{\|})$ in the right 
Moyal-Rindler wedge, defined as 
\begin{equation}
\Phi_{\theta}^{R,1}(t,x_{\|})= \left(\Phi_{0}^{R,1(+)}(t,x_{\|}) + \Phi_{0}^{R,1(-)}(t,x_{\|})\right)
e^{\frac{1}{2}\overleftarrow{\partial}\wedge\mathcal{P}} \equiv 
\Phi_{\theta}^{R,1(+)}(t,x_{\|})+\Phi_{\theta}^{R,1(-)}(t,x_{\|}),
\end{equation}
 where $\Phi_{0}^{R,1(+)}(t,x_{\|})$ and $\Phi_{0}^{R,1(-)}(t,x_{\|})$ are the annihilation and creation 
 parts respectively of $\Phi_{0}^{R,1}(t,x_{\|})$. 
The annihilation and creation parts of $\Phi_{\theta}^{R,1}(t,x_{\|})$ are 
defined as 
\begin{eqnarray}
 \Phi_{\theta}^{R,1(\pm)}(t,x_{\|}) \equiv \Phi_{0}^{R,1(\pm)}(t,x_{\|})e^{\frac{1}{2}\overleftarrow{\partial}_{\mu}\theta^{\mu\nu}\mathcal{P}_{\nu}}.
\end{eqnarray}
Simple computation using (\ref{k+w},\ref{-k+w}) and (\ref{global_eigen3}) shows that the exponent of 
the twist factor appearing in $\Phi_{\theta}^{R,1}(t,x_{\|}$) is proportional to the particular operator 
combination $(\mathcal{P}_0+\mathcal{P}_1)$. 
Similarly, the exponent of the twist factor in $\Phi_{\theta}^{R,2}(t,x_{\|})$ is proportional to $(\mathcal{P}_0-\mathcal{P}_1)$. 
This fact is to be noted as it will be needed for discussions later.

Now let us see what happens when we twist the massive field in $(1+1)$ dimension. 
 We will again use (\ref{twist_rindler}) to twist the quantum field,
with $\Phi_{0}^{R}(t,x_{\|})$ given by equation (\ref{rind-minkcoord}). 
The massive twisted field $\Phi_{\theta}^{R}(t,x_{\|})$ in $(1+1)$ dimensions is given by
\begin{equation}
\Phi_{\theta}^{R}(t,x_{\|})=\int_{0}^{\infty}\frac{d\omega}{(2\pi)(2\omega)} 
\left(a_{\omega}^{R}\varphi_{\omega,R}(t,x_{\|}) 
e^{\frac{1}{2}\overleftarrow{\partial}\wedge \mathcal{P}} + 
a_{\omega}^{R \dagger} \varphi_{\omega,R}^{*}(t,x_{\|}) 
e^{\frac{1}{2}\overleftarrow{\partial}\wedge\mathcal{P}} \right)
\label{twisted-rindler-qft}
\end{equation}
where $\varphi_{\omega,R}(t,x_{\|})$ is given by (\ref{eigenfunc})
and $a_{\omega}^{R},a_{\omega}^{R \dagger}$ satisfy (\ref{rind-commutator5}).
Equivalently,
\begin{equation}
 \Phi_{\theta}^{R}(t,x_{\|})=\int_{0}^{\infty}\frac{d\omega}{(2\pi)(2\omega)} 
 \left(\phi_{\omega,\theta}^{R (+)} (t,x_{\|})+\phi_{\omega,\theta}^{R (-)} (t,x_{\|}) \right),
\end{equation}
where $\phi_{\omega,\theta}^{R (+)} (t,x_{\|})= a_{\omega}^{R}\varphi_{\omega,R}(t,x_{\|})
e^{\frac{1}{2}\overleftarrow{\partial}\wedge \mathcal{P}}$ is the annihilation part of 
twisted field with mode $\omega$, and  $\phi_{\omega,\theta}^{R (-)} 
(t,x_{\|})= a_{\omega}^{R\dagger}\varphi_{\omega,R}^{\ast}(t,x_{\|})e^{\frac{1}{2}\overleftarrow{\partial}\wedge \mathcal{P}}$ is the corresponding creation part.
 

Finally, the twisted quantum field in $(3+1)$ dimensions, using (\ref{rind-minkcoord2}) and (\ref{twist_rindler}), is given by
\begin{equation}
 \Phi_{\theta}^{R}(t,x_{\|},\vec{x}_\perp) = \int\frac{d^{2}\vec{k}_{\perp}d\omega}{(2\pi)^3 (2\omega)}
 \left(a_{\omega,\vec{k}_\perp}^{R} \varphi_{\omega,\vec{k}_\perp,R}(t,x_{\|},\vec{x}_\perp)
 e^{\frac{1}{2}\overleftarrow{\partial}\wedge \mathcal{P}} + a_{\omega,\vec{k}_\perp}^{R \dagger} 
 \varphi_{\omega,\vec{k}_\perp,R}^{*}(t,x_{\|},\vec{x}_\perp) 
 e^{\frac{1}{2}\overleftarrow{\partial}\wedge \mathcal{P}}\right)
\label{twisted-rindler-qft2}
\end{equation}
where $\varphi_{\omega,\vec{k}_\perp,R}(t,x_{\|},\vec{x}_\perp)$ is given by (\ref{eigenfunc2}) and the operators $a_{\omega,\vec{k}_\perp}^{R}$ and 
$a_{\omega,\vec{k}_\perp}^{R \dagger}$ satisfy (\ref{rind-commutator5}). 

%
%
\subsection*{Wedge localization of properties $\Phi_{\theta}^{R}$}
One of the the most important properties of quantum field in Rindler wedges is wedge localization -- 
the quantum field $\Phi_{0}^{R}$ (or $\Phi_{0}^{L}$) is localized in the right (left) Rindler wedge 
and vanishes in the other Rindler wedge. Not only does $\Phi_{0}^{R}$ (or $\Phi_{0}^{L}$) vanishes in 
$L (R)$ but it also vanishes on the lightcone : 
\begin{eqnarray}
\Phi_{0}^{R}(t,x_{\|},\vec{x}_\perp)&=&0 \quad \forall \quad (t,x_{\|},\vec{x}_\perp) \in L \quad {\rm and} \quad t\pm x_{\|}=0, \\
\Phi_{0}^{L}(t,x_{\|},\vec{x}_\perp)&=&0 \quad \forall \quad (t,x_{\|},\vec{x}_\perp) \in R \quad {\rm and} \quad  t\pm x_{\|}=0.
\end{eqnarray}
This is precisely due to the wedge localization property of the functions $\varphi_{\omega,\vec{k}_\perp,R}$ and 
$\varphi_{\omega,\vec{k}_\perp,L}$, as is seen from (\ref{eigenkrd}) and (\ref{eigenkld}).

Is the twisted quantum field $ \Phi_{\theta}^{R}(t,x_{\|},\vec{x}_\perp)$ wedge localized? 
$\Phi_{\theta}^{R}(t,x_{\|},\vec{x}_\perp)$ contains $\varphi_{\omega,R}(t,x_{\|},\vec{x}_\perp)$ along 
with its derivatives to all orders. So though $\varphi_{\omega,R}(t,x_{\|},\vec{x}_\perp)$ goes to zero 
in the left Rindler wedge and on the light cone, there is nothing that guarantees the vanishing of all 
the derivatives in these regions. It is easily seen that all the derivatives vanish in the left Rindler 
wedge but do not go to zero on the lightcone. So
 \begin{eqnarray}
\Phi_{\theta}^{R (L)}(t,x_{\|},\vec{x}_\perp) \left\{
\begin{array}{l l l}
=0 & \forall & (t,x_{\|},\vec{x}_\perp) \in L (R),\\
\neq 0 & \forall & t\pm x_{\|}=0.
\end{array} \right.
\end{eqnarray}

The twisted quantum field $\Phi_{\theta}^{R}(t,x_{\|},\vec{x}_\perp)$ and  
$\Phi_{\theta}^{L}(t,x_{\|},\vec{x}_\perp)$  are wedge localized as they vanish in $L$ and $R$ (in 
the sense above), but their properties on the lightcone are different than their commutative 
counterparts.

\subsection*{Hermiticity}
The quantum fields $\Phi_{0}^{R}$ and $\Phi_{0}^{L}$ are hermitian: $\Phi_{0}^{R(L)\dagger} = 
\Phi_{0}^{R(L)}$. What can one say about the hermiticity of the twisted quantum fields 
$\Phi_{\theta}^{R}$ and $\Phi_{\theta}^{L}$?

First let us consider massless case in $(1+1)$ dimension. The creation/annihilation operators on 
the Rindler wedge are related to those in Minkowski spacetime by the Bogoliubov transformation 
(\ref{bogolibov1}). As we saw earlier, the twist factor for the right-moving  part of the massless 
field involves $\mathcal{P}_{0}+\mathcal{P}_{1}$. From equations (\ref{bogolibov1}) and 
(\ref{field_mom1}), it is easily seen that 
\begin{eqnarray}
[a_{\omega,1}^{R},\mathcal{P}_{0}+\mathcal{P}_{1}]=0.
\label{alg_1}
\end{eqnarray}
Hence  $(\Phi_{\theta}^{R,1(+)})^{\dagger}=\Phi_{\theta}^{R,1(-)}$ which implies that 
$\Phi_{\theta}^{R,1}$ is hermitian. Similarly,
\begin{eqnarray}
[a_{\omega,2}^{R},\mathcal{P}_{0}-\mathcal{P}_{1}]=0.
\label{alg_2}
\end{eqnarray}
So $\Phi_{\theta}^{R,2}$ and therefore 
$\Phi_{\theta}^{R(L)}$ is hermitian. Hermiticity of the quantum field is thus unaffected by twisting.
$\mathcal{P}_{0}\pm\mathcal{P}_{1}$ forms the central element of the algebras (\ref{alg_1}) and 
(\ref{alg_2})  and what is very striking is that the twist  conspires to produce that central element 
in exponential.

Next let us look at the $(3+1)$-dimensional (both massless and massless) case. 
The exponent in twist element takes the form $\partial_{\mu}\theta^{\mu\nu}\mathcal{P}_{\nu} = 
\partial_{t}\theta^{0i}\mathcal{P}_{i}+\partial_{x^{i}}\theta^{i0}\mathcal{P}_{0} + 
\partial_{x^{i}}\theta^{ij}\mathcal{P}_{j}$ for $ i,j=1,2,3$. The most general $\theta$-matrix can be 
written as
\begin{eqnarray}
 \theta=\left[ \begin{array}{cccc}
0 & \theta^{01} & \theta^{02} & \theta^{03} \\
\theta^{10}&0 & \theta^{12} &\theta^{13} \\
\theta^{20}& \theta^{21} & 0& \theta^{23} \\
\theta^{30}&\theta^{31} & \theta^{32} & 0\end{array} \right]\equiv\left[ \begin{array}{cccc}
0 & -E^{1} & -E^{2} & -E^{3} \\
E^{1}& 0 & -B^{3} & B^{2} \\
E^{2}& B^{3} & 0& -B^{1}  \\
E^{3}& -B^{2} & B^{1} & 0\end{array} \right].
\label{gen_thetamatrix}
\end{eqnarray}
The exponent of the twist element simplifies to 
$(-\partial_{t}(\vec{E}\cdot \vec{\mathcal{P}})+(\vec{\nabla}_{\vec{x}}\cdot \vec{E})\mathcal{P}_0 + 
\vec{\nabla}_{\vec{x}}\cdot(\vec{B}\times\vec{\mathcal{P}}))$. 
Since $a_{\omega,\vec{k}_\perp}^{R}$  and $b_{\vec{k}}$ are related by (\ref{bogolibov2}),
it is easy to see that in general $a_{\omega,\vec{k}_\perp}^{R(L)}$ and 
$a_{\omega,\vec{k}_\perp}^{R(L) \dagger}$ do not commute with $\mathcal{P}_{\mu}$: 
\begin{equation}
 [a_{\omega,\vec{k}_\perp}^{R},\mathcal{P}_{\mu}] \neq 0 \quad \mathrm{and} \quad 
 [a_{\omega,\vec{k}_\perp}^{R \dagger},\mathcal{P}_{\mu}] \neq 0 .
\label{twist-awR-commutation}
\end{equation}
So the above twist factor in general does not commute with the $a_{\omega,\vec{k}_\perp}^{R}$ 
and $(\phi_{\omega,\vec{k}_\perp,\theta}^{R (+)})^{\dagger} \neq \phi_{\omega,,\vec{k}_\perp,\theta}^{R (-)}$. 
Hence the twisted quantum field $\Phi_{\theta}^{R}(t,x_{\|},\vec{x}_\perp)$ is not hermitian. 


The discussion for the massive case in $(1+1)$ dimensions is conceptually identical to the one above: 
again, the twisted massive quantum field is not hermitian.

\section{Quantum Correlation Functions}

In commutative spacetime, the Minkowski vacuum  is a KMS thermal state at temperature 
$T=\frac{a}{2\pi}$ for an accelerating observer. The accelerating observer detects a Bose-Einstein  
distribution with this temperature, which may be explicitly derived from the Bogoliubov coefficients in 
(\ref{bogolibov2}) \cite{crispino,takagi}.

It  is well known in  algebraic quantum field theory that a state $|\psi\rangle$ is a KMS thermal state at 
temperature $T=\frac{1}{\beta}$ if there exists an 
operator $\hat{J}$ (which commutes with the time translation operator:  
$\hat{J} e^{-i\Hat{H}\tau}=e^{-i\hat{H}\tau}\hat{J} \quad \forall \tau \in {\mathbb R}$)
such that
\begin{equation}
 e^{-\hat{H}\beta/2} \hat{A} |\psi\rangle = \hat{J} \hat{A}^{\dagger}|\psi\rangle
\label{biwi}
\end{equation}
where $\hat{A}$ is an element of the algebra of observables. 
This is the Bisognano-Wichmann theorem \cite{bisognano} (see also the discussion in 
\cite{sewell}). 

If $e^{-i\hat{\mathcal{K}}\alpha}$ is the boost that generates the transformation
\begin{eqnarray}
 && t\rightarrow t'=t\cosh (a\alpha) +x_{\|} \sinh (a\alpha), \nonumber \\
&& x_{\|} \rightarrow x_{\|}'=t \sinh (a\alpha) + \cosh (a\alpha), \nonumber \\
&&\vec{x}_{\perp}\rightarrow \vec{x'}_{\perp}=\vec{x}_{\perp}, \nonumber
\end{eqnarray}
then for $\alpha=i\pi/a$, we get $t\rightarrow -t$, $x_{\|}\rightarrow -x_{\|}$ and 
$\vec{x}_\perp \rightarrow \vec{x}_{\perp}$. This can be seen from 
(\ref{finite_trans}--\ref{finite_trans1}) of the appendix. It is thus straightforward to show for any 
$N \geq 1$
\begin{eqnarray}
&& e^{-\hat{\mathcal{K}}\pi/a}\prod_{n=1}^{N}\Phi_{0}^{R}(t_{n},x_{\| n},\vec{x}_{\perp n})|0_{M}\rangle=\prod_{n=1}^{N}\Phi_{0}^{R}(-t_{n},-x_{\| n},\vec{x}_{\perp n})|0_{M}\rangle
\label{biwi2}
\end{eqnarray}
 as $\hat{\mathcal{K}}|0_{M}\rangle=0$. Equivalently, 
\begin{eqnarray}
&& e^{-\hat{\mathcal{K}}\pi/a}\prod_{n=1}^{N}\Phi_{0}^{R}(t_{n},x_{\| n},\vec{x}_{\perp n})|0_{M}\rangle= \hat{J}\prod_{n=1}^{N}\Phi_{0}^{R}(t_{n},x_{\| n},\vec{x}_{\perp n})|0_{M}\rangle,
\label{biwi3}
\end{eqnarray}
where $\hat{J}$ is the $PCT$ times $\pi$- rotation about $\hat{n}$. (In $(1+1)$ dimensions, $\hat{J}$ 
is just the $PCT$ operator). This equation is of the same form as (\ref{biwi}) with $$\hat{A} = 
\prod_{n=1}^{N} \phi_{0}^{R}(t_{n},x_{\| n},\vec{x}_{\perp n})$$ for any $ N \geq 1$. 
Hence $|0_{M}\rangle$ is a KMS thermal state at $T=\frac{a}{2\pi}$. 

Let us consider the massless field in $(1+1)$ dimension. In commutative case, the  massless field 
in either Rindler wedge satisfies the Bisognano-Wichmann theorem (\ref{biwi3}) (with $\hat{J}$ as 
the $PCT$ operator). Alternately, from  (\ref{massless_field_RRW}) and (\ref{bogolibov1}) we can 
show that  
 \begin{eqnarray}
\langle0_{M} |a_{\omega,\lambda}^{R \dagger}a_{\omega^{\prime},\lambda^{\prime}}^{R}|0_{M}\rangle &=& (2\pi)
(2\omega)\delta_{\lambda \lambda^{\prime}}\frac{\delta(\omega-\omega^{\prime})}{e^{2\pi\omega/a}-1}
\label{B_E1}
 \end{eqnarray}
which is just the B-E distribution at $T=\frac{a}{2\pi}$.  
The same distribution can be obtained from the two point correlation function  
$\langle0_{M} |\Phi_{0}^{R (-)} (t,x_{\|}) \Phi_{0}^{R (+)} (t^{\prime},
x_{\|}^{\prime})|0_{M}\rangle$ using the Bisognano-Wichmann theorem \cite{crispino}.

In the GM plane, as the massless field in Moyal Rindler wedge is hermitian, it satisfies (\ref{biwi3}) with 
$\hat{A}=\Phi_{\theta}^{R}(t,x_{\|})$ and $\hat{J}=PCT$: 
\begin{equation}
e^{-\hat{\mathcal{K}}\pi/a}
\Phi_{\theta}^{R}(t,x_{\|})|0_{M}\rangle= \hat{J}\Phi_{\theta}^{R}(t,x_{\|})|0_{M}\rangle. 
\end{equation}
But if we consider a string of more than one field operators i.e $\hat{A}=\prod_{i=1}^{N}\Phi_
{\theta}^{R}(t^{i},x^{i}_{\|})$ for $N\geq 2$, then (\ref{biwi3}) is not satisfied. As the elements of the 
operator algebra does not satisfy (\ref{biwi3}), Bisognano-Wichmann theorem is not satisfied. So  
$|0_M\rangle $, unlike the commutative case, is not a KMS thermal state.

Now let us find out the noncommutative (Moyal) analogue of the two point correlation function
 $\langle0_{M} |\Phi_{\theta}^{R (-)} (t,x_{\|})\Phi_{\theta}^{R (+)} (t^{\prime},x_{\|}^{\prime})
 |0_{M}\rangle$. Using (\ref{twist_rindler}), the Moyal two-point correlation function can be written as
\begin{equation}
 \langle0_{M} |\Phi_{\theta}^{R (-)} (t,x_{\|}) \Phi_{\theta}^{R (+)} (t^{\prime},x_{\|}^{\prime})|0_{M}\rangle= 
\langle0_{M} |\Phi_{0}^{R (-)} (t,x_{\|})e^{\frac{\tilde{\theta}}{2}(\overleftarrow{\partial}_{t}\mathcal{P}_{1}-\overleftarrow{\partial}_{x_{\|}}\mathcal{P}_{0})}  
\Phi_{0}^{R (+)} (t^{\prime},x_{\|}^{\prime}) 
e^{\frac{\theta}{2}(\overleftarrow{\partial}_{t^{\prime}}\mathcal{P}_{1}-\overleftarrow{\partial}_{x_{\|}^{\prime}}\mathcal{P}_{0})}|0_{M}\rangle . 
\label{B-E-theta6} 
\end{equation}
But $a_{\omega,\lambda}^{R}$ commutes with the twist factor and 
\begin{equation}
 \mathcal{P}_{\mu}|0_{M}\rangle = 0.
\label{p_mu_vacuum}
\end{equation}
So the noncommutative two-point correlation function reduces to 
its commutative counterpart:
$ \langle0_{M} |\Phi_{\theta}^{R (-)}(t,x_{\|}) \Phi_{\theta}^
{R (+)} (t^{\prime},x_{\|}^{\prime})|0_{M}\rangle=\langle0_{M}
 |\Phi_{0}^{R (-)} (t,x_{\|}) \Phi_{0}^{R (+)} (t^{\prime},x_{\|}
^{\prime})|0_{M}\rangle$ and hence carries 
no information about noncommutativity.

Let us now consider the four-point correlation function 
$$\langle0_{M}|\Phi_{\theta}^{R (-)}(t^1,x^1_{\|})\Phi_{\theta}^{R (-)}(t^2,x^2_{\|})\Phi_{\theta}^
{R (+)} (t^{3},x_{\|}^{3})\Phi_{\theta}^
{R (+)} (t^{4},x_{\|}^{4})|0_{M} \rangle.$$
 Owing to the relation
\begin{equation}
 [\Phi_{0}^{R}(t,x_{\|}), \mathcal{P}_{\mu}]=i\partial_{\mu}\Phi_{0}^{R}(t,x_{\|})
\label{p_del}
\end{equation} 
and using (\ref{p_mu_vacuum}), the four-point correlation function simplifies to 
\begin{equation}
 \langle0_{M}|\Phi_{0}^{R (-)}(t^1,x^1_{\|})e^{\frac{i}{2}\overleftarrow{\partial}_{1}\wedge\overrightarrow{\partial}_{2}}\Phi_{0}^{R (-)}(t^2,x^2_{\|})\Phi_{0}^
{R (+)} (t^{3},x_{\|}^{3})e^{-\frac{i}{2}\overleftarrow{\partial}_{3}\wedge\overrightarrow{\partial}_{4}}\Phi_{0}^
{R (+)} (t^{4},x_{\|}^{4})|0_{M} \rangle.
\end{equation}
As all the derivatives do not vanish, the four-point correlation function (as well as all higher order ones) 
deviate from the commutative result. 


Next let us discuss massive case in $(1+1)$ dimension. In commutative case, 
(\ref{biwi3}) is satisfied for massive fields as well. So the state $|0_M\rangle$ 
is a KMS thermal state at $T=\frac{a}{2\pi}$. The distribution function can again be computed from 
the Bogoluibov coefficients as
  \begin{eqnarray}
\langle0_{M} |a_{\omega}^{R \dagger}a_{\omega^{\prime}}^{R}|0_{M}\rangle &=& (2\pi)(2\omega)\frac{\delta(\omega-\omega^{\prime})}{e^{2\pi\omega/a}-1},
 \label{B_E2}
\end{eqnarray} 
which is that same as obtained from the two-point function $\langle 0_M |\Phi_{0}^{R -}(t,x_{\|})\Phi_{0}^{R +}(t^{\prime},x_{\|}^{\prime})|0_M \rangle$.

In Moyal Rindler wedge, due to the non-hermiticity of the massive quantum fields, Bisognano Wichmann 
theorem is not satisfied because (\ref{biwi3}) does not hold: $|0_{M}\rangle$ is not an equilibrium 
thermal state. Nonetheless, we can still compute the two-point correlation function and find the 
deviation from the commutative case. Using (\ref{twist_rindler}), this can be written as 
\begin{equation}
 \langle0_{M} |\Phi_{\theta}^{R (-)} (t,x_{\|}) \Phi_{\theta}^{R (+)} (t^{\prime},x_{\|}^{\prime})|0_{M}\rangle= 
\langle0_{M} |\Phi_{0}^{R (-)} (t,x_{\|})e^{\frac{\tilde{\theta}}{2}(\overleftarrow{\partial}_{t}\mathcal{P}_{1}-\overleftarrow{\partial}_{x_{\|}}\mathcal{P}_{0})}  
\Phi_{0}^{R (+)} (t^{\prime},x_{\|}^{\prime})
e^{\frac{\theta}{2}(\overleftarrow{\partial}_{t^{\prime}}\mathcal{P}_{1}-\overleftarrow{\partial}_{x_{\|}^{\prime}}\mathcal{P}_{0})}|0_{M}\rangle. 
\label{B-E-theta7} 
\end{equation}


Using  (\ref{p_mu_vacuum}) and (\ref{p_del}), the two-point correlation function can be simplified as
\begin{eqnarray}
 \langle0_{M} |\Phi_{\theta}^{R (-)} (t,x_{\|}) \Phi_{\theta}^{R (+)} (t^{\prime},x^{\prime}_{\|})|0_{M}\rangle &= \langle0_{M} |\Phi_{0}^{R (-)} (t,x_{\|}) e^{-i\frac{\theta}{2}(\overleftarrow{\partial}_{t}\overrightarrow{\partial}_{x_{\|}{\prime}}-\overleftarrow{\partial}_{x_{\|}}\overrightarrow{\partial}_{t^{\prime}})}\Phi_{0}^{R (+)} (t^{\prime},x^{\prime}_{\|})|0_{M}\rangle\\
&=e^{-i\frac{\theta}{2}(\partial_{t}\partial_{x_{\|}{\prime}}-\partial_{x_{\|}}\partial_{t^{\prime}})}\langle0_{M} |\Phi_{0}^{R (-)} (t,x_{\|}) \Phi_{0}^{R (+)} (t^{\prime},x^{\prime}_{\|})|0_{M}\rangle.
\end{eqnarray}
Substituting from (\ref{rind-minkcoord}), we see that unlike the massless case in $(1+1)$ dimension, 
here the two-point correlation function has also changed. The deviation from the commutative result is 
\begin{eqnarray}
 \left( e^{-i\frac{\theta}{2}(\partial_{t}\partial_{x_{\|}{\prime}}-\partial_{x_{\|}}\partial_{t^{\prime}})}-1\right)\langle0_{M} |\Phi_{0}^{R (-)} (t,x_{\|}) \Phi_{0}^{R (+)} (t^{\prime},x^{\prime}_{\|})|0_{M}\rangle.
\end{eqnarray}
%
%
%

In $(3+1)$ dimensional commutative spacetime, the massive hermitian quantum fields satify (\ref{biwi3}). So $|0_M\rangle$ is a 
thermal state and the number distribution function is 
\begin{eqnarray}
\langle0_{M} |a_{\omega,\vec{k}_\perp}^{R \dagger}a_{\omega^{\prime},\vec{k}_\perp^\prime}^{R}|0_{M}\rangle &=& (2\pi)^3(2\omega)\frac{\delta(\omega-\omega^{\prime})}{e^{2\pi\omega/a}-1}\delta^{2}(\vec{k}_\perp-\vec{k}_{\perp}^{\prime})
 \label{B_E3}
\end{eqnarray}
and the two-point correlation function is $\langle0_{M} |\Phi_{0}^{R (-)} (t,x_{\|},\vec{x}_{\perp}) \Phi_{0}^{R (+)} (t,x_{\|},
\vec{x}_{\perp})|0_{M}\rangle$  is
\begin{eqnarray}
\langle0_{M} |\Phi_{0}^{R (-)} (t,x_{\|},\vec{x}_{\perp}) \Phi_{0}^{R (+)} (t,x_{\|},\vec{x}_{\perp})|0_{M}\rangle=\int \frac{d\omega d^{2}\vec{k}_{\perp}}{(2\pi)^{3}(2\omega)}\frac{d\omega^\prime d^{2}\vec{k}_{\perp}^{\prime}}{(2\pi)(2\omega^\prime)}
(2\pi)^{3}(2\omega) \nonumber\\
\frac{\delta(\omega-\omega^{\prime})\delta^{2}(\vec{k}_\perp-\vec{k}_{\perp}^{\prime})}{e^{2\pi\omega/a}-1} \varphi_{\omega,\vec{k}_{\perp},R}^{*}(t,x_{\|},\vec{x}_{\perp})\varphi_{\omega^\prime,\vec{k}_{\perp}^{\prime},R}(t,x_{\|},\vec{x}_{\perp}).
\label{2pt-massive2}
\end{eqnarray}

The twisted fields in Rindler wedges do not satisfy a relation like (\ref{biwi3}). Although $|0_M\rangle$ is not a state in thermal 
equilibrium, we can compute the two-point correlation function in the Moyal-Rindler wedge: 
\begin{equation}
 \langle0_{M} |\Phi_{\theta}^{R (-)} (t,x_{\|},\vec{x}_{\perp}) \Phi_{\theta}^{R (+)} (t',x'_{\|},\vec{x}'_{\perp})|0_{M}\rangle=
\langle0_{M} |\Phi_{0}^{R (-)} (t,x_{\|},\vec{x}_{\perp})e^{\frac{1}{2}\overleftarrow{\partial}_{x}\wedge\mathcal{P}}
\Phi_{0}^{R (+)} (t^{\prime},x_{\|}^{\prime},\vec{x}_{\perp}^{\prime}) 
e^{\frac{1}{2}\overleftarrow{\partial}_{x^{\prime}}\wedge\mathcal{P}}|0_{M}\rangle.
\end{equation}
Using (\ref{p_mu_vacuum}) and (\ref{p_del}), we can express the the correlation function as 
\begin{equation}
 \langle0_{M} |\Phi_{\theta}^{R (-)} (t,x_{\|},\vec{x}_{\perp}) \Phi_{\theta}^{R (+)} (t',x'_{\|},\vec{x}'_{\perp})|0_{M}\rangle
= e^{-\frac{i}{2}\overleftarrow{\partial}_{x}\wedge\overrightarrow{\partial}_{x^{\prime}}}\langle0_{M} |\Phi_{0}^{R (-)} (t,x_{\|},\vec{x}_{\perp})\Phi_{0}^{R (+)} (t^{\prime},x_{\|}^{\prime},\vec{x}_{\perp}^{\prime})|0_{M}\rangle.
\end{equation}
Again this is different from its commutative counterpart, as all the derivatives of the commutative 
two-point correlation function do not vanish. The deviation from the commutative case is  
\begin{eqnarray}
\left(e^{-\frac{i}{2}\overleftarrow{\partial}_{x}\wedge\overrightarrow{\partial}_{x^{\prime}}}-1\right)\langle0_{M} |\Phi_{0}^{R (-)} (t,x_{\|},\vec{x}_{\perp})\Phi_{0}^{R (+)} (t^{\prime},x_{\|}^{\prime},\vec{x}_{\perp}^{\prime})|0_{M}\rangle .
\label{deviation}
\end{eqnarray}

\subsection{Arbitrary $\theta^{\mu\nu}$}
The most general $\theta$ matrix is given in equation (\ref{gen_thetamatrix}). 
The exponential factor $\partial_{x}\wedge\partial_{x^{\prime}}$ in (\ref{deviation}) can be simplified by recognizing that the four vectors 
$\vec{E}, \vec{B}, \vec{x}$ and $\vec{x^{\prime}}$ can be decomposed into components that are parallel and perpendicular to $\hat{n}$. With 
this decomposition we get
\begin{eqnarray}
\overleftarrow{\partial}_{x}\wedge\overrightarrow{\partial}_{x^{\prime}}=&-\overleftarrow{\partial}_{t}((\vec{E}\cdot\hat{n}) 
(\hat{n}\cdot\vec{\nabla}_{\vec{x}^{\prime}})+\vec{E}_{\perp}\cdot\vec{\nabla}_{\vec{x}_\perp^{\prime}} )+((\hat{n}\cdot
\overleftarrow{\nabla}_{\vec{x}})(\vec{E}\cdot\hat{n})+\overleftarrow{\nabla}_{\vec{x}_\perp}\cdot \vec{E}_\perp)
\overrightarrow{\partial}_{t^{\prime}} + \overleftarrow{\nabla}_{\vec{x}_\perp}\cdot (\nonumber \\
& \vec{B}_\perp\times\hat{n}(\hat{n}\cdot\vec{\nabla}_{\vec{x}^{\prime}} ))+\overleftarrow{\nabla}_{\vec{x}_\perp}\cdot
((\vec{B}\cdot\hat{n}) \hat{n}\times\vec{\nabla}_{\vec{x}_\perp^{\prime}})+\hat{n}(\hat{n}\cdot\overleftarrow{\nabla}_{\vec{x}})
\cdot(\vec{B}_\perp\times\vec{\nabla}_{\vec{x}_\perp^{\prime}}).
\end{eqnarray}
For special choices of the $\vec{E}$ and $\vec{B}$, the above expression simplifies. 
For example if $\vec{E}=0=\vec{B}$, then we get back the commutative spacetime, and the two-point correlation function reduces to its commutative counterpart
as $\overleftarrow{\partial}_{x}\wedge\overrightarrow{\partial}_{x^{\prime}} =0$.

If we choose $\vec{E}=0,\vec{B} \neq 0 $, (magnetic-type noncommutativity), then the exponential factor
\begin{equation}
  \overleftarrow{\partial}_{x}\wedge\overrightarrow{\partial}_{x^{\prime}}=  \overleftarrow{\nabla}_{\vec{x}_\perp}\cdot 
  (\vec{B}_\perp\times\hat{n}(\hat{n}\cdot\vec{\nabla}_{\vec{x}^{\prime}} ))+\overleftarrow{\nabla}_{\vec{x}_\perp}\cdot
  ((\vec{B}\cdot\hat{n}) \hat{n}\times\vec{\nabla}_{\vec{x}_\perp^{\prime}})+\hat{n}(\hat{n}\cdot
  \overleftarrow{\nabla}_{\vec{x}})\cdot(\vec{B}_\perp\times\vec{\nabla}_{\vec{x}_\perp^{\prime}})
\end{equation}
involves only spatial derivatives.
On the other hand, if we chose $\vec{E} \neq 0, \vec{B} = 0 $ (electric-type noncommutativity), then  
\begin{equation}
\overleftarrow{\partial}_{x}\wedge\overrightarrow{\partial}_{x^{\prime}}=-\overleftarrow{\partial}_{t}((\vec{E}\cdot\hat{n}) 
(\hat{n}\cdot\vec{\nabla}_{\vec{x}^{\prime}})+\vec{E}_{\perp}\cdot\vec{\nabla}_{\vec{x}_\perp^{\prime}} )+((\hat{n}\cdot
\overleftarrow{\nabla}_{\vec{x}})(\vec{E}\cdot\hat{n})+\overleftarrow{\nabla}_{\vec{x}_\perp}\cdot \vec{E}_\perp)
\overrightarrow{\partial}_{t^{\prime}}.
\end{equation}
We can also consider light-like noncommutativity $\vec{E}=\vec{B}$. If $\vec{E}_\perp=\vec{B}_\perp = 0, E_\|=B_\| $, then 
\begin{eqnarray}
 \overleftarrow{\partial}_{x}\wedge\overrightarrow{\partial}_{x^{\prime}}=&-\overleftarrow{\partial}_{t}(\vec{E}\cdot\hat{n}) 
 (\hat{n}\cdot\vec{\nabla}_{\vec{x}^{\prime}})+(\hat{n}\cdot\overleftarrow{\nabla}_{\vec{x}})(\vec{E}\cdot\hat{n})
 \overrightarrow{\partial}_{t^{\prime}}+\overleftarrow{\nabla}_{\vec{x}_\perp}\cdot((\vec{B}\cdot\hat{n}) 
 \hat{n}\times\vec{\nabla}_{\vec{x}_\perp^{\prime}}),
\end{eqnarray}
while if choose $\vec{E}_\perp=\vec{B}_\perp , E_\|=B_\|=0 $, then
\begin{eqnarray}
\overleftarrow{\partial}_{x}\wedge\overrightarrow{\partial}_{x^{\prime}}=-\overleftarrow{\partial}_{t}
\vec{E}_{\perp}\cdot\vec{\nabla}_{\vec{x}_\perp^{\prime}} +\overleftarrow{\nabla}_{\vec{x}_\perp}\cdot \vec{E}_\perp
\overrightarrow{\partial}_{t^{\prime}} + \overleftarrow{\nabla}_{\vec{x}_\perp}\cdot (
 \vec{B}_\perp\times\hat{n}(\hat{n}\cdot\vec{\nabla}_{\vec{x}^{\prime}} ))+\nonumber \\
\hat{n}(\hat{n}\cdot\overleftarrow{\nabla}_{\vec{x}})
\cdot(\vec{B}_\perp\times\vec{\nabla}_{\vec{x}_\perp^{\prime}}).
\end{eqnarray}

The maximum simplification can happen if one chooses the $\vec{E}$ along $\hat{n}$ i.e. 
$\vec{E} = E_{\|}\hat{n}$ and $\vec{B} =0$. Then the exponential factor reduces to  
\begin{eqnarray}
\partial_{x}\wedge\partial_{x^{\prime}}&=&-E_{\|}(\overleftarrow{\partial}_{t} 
(\hat{n}\cdot\vec{\nabla}_{\vec{x}^{\prime}}) -
(\hat{n}\cdot\vec{\nabla}_{\vec{x}})\overrightarrow{\partial}_{t^{\prime}})
\end{eqnarray}
which is same as the result we get in $(1+1)$ dimensional case.

Thus effect of noncommutativity persists and is always carried by two-point correlation function, 
unless we take both $\vec{E}=0$ and $\vec{B}=0$. 

\section{Conclusion}

The standard Unruh effect is fascinating because it captures certain generic features of quantum 
field theories on spacetimes that have causal horizons. It succinctly encapsulates the physics that 
underlies Hawking radiation from black holes.

Our analysis shows that while it is indeed possible to rigorously define an accelerating observer in 
GM spacetime, there are new and interesting effects in this case. The Minkowski vacuum is no longer 
a thermal state for the accelerating observer, and the ``Rindler'' particles do not obey 
Bose-Einstein distribution. 

It is tempting to use the lessons of this analysis to speculate on the implications for the physics of 
black holes in noncommutative spacetimes. In particular, it would be interesting to investigate in detail 
the connection between black holes and thermodynamics, and that of black hole evaporation. 
Questions like these can perhaps be answered only by doing detailed computations in the context 
of specific models involving black holes in noncommutative spacetimes, something we leave for 
future investigation.

{\bf Acknowledgments:} It is a pleasure to thank A. P. Balachandran for many illuminating discussions, 
and for pointing out an error during the early stages of this project.

\section*{Appendices}

\appendix

\section{Eigenfunctions of the Boost Operator \label{e_b_o}}

In $d+1$ dimensions, the eigenvalues and eigenfunctions of generator of a Lorentz boost along an arbitrary direction 
$\hat{n}$ are obtained by solving the equation
\begin{equation}
\mathcal{K} \varphi_{\omega,\vec{k}_\perp}(t,x_{\|}, \vec{x}_{\perp}) = 
 \omega\varphi_{\omega, \vec{k}_\perp}(t,x_{\|},\vec{x}_{\perp}), \quad \mathcal{K} \equiv   
  \vec{\mathcal{K}} \cdot \hat{n}, \quad  -\infty<\omega<\infty,
\label{boost_eve1} 
\end{equation}
where  $\omega$ is a continuous real eigenvalue and $\vec{k}_\perp$ is the $d-1$ dimensional vector component of spatial 
momentum $\vec{k}=(k_\|,\vec{k}_\perp)$.
These eigenfunctions form a complete orthonormal set \cite{hughes}
\begin{eqnarray}
 \langle\varphi_{\omega,\vec{k}_\perp}(t,x_{\|},\vec{x}_{\perp})|\varphi_{\omega^{\prime},\vec{k}_\perp^{\prime}}(t,x_{\|},\vec{x}_{\perp})\rangle_{M} &=& \delta(\omega-\omega^{\prime})\delta^{d-1}(\vec{k}_\perp-\vec{k}_\perp^{\prime}).
\label{orthonormal5}
\end{eqnarray}

It is easy to show that
\begin{equation}
 \varphi_{\omega,\vec{k}_\perp}(t\cosh a\alpha + x_{\|}\sinh a\alpha,t\sinh a\alpha + 
 x_{\|}\cosh a\alpha,\vec{x}_{\perp}) = e^{-i\omega a\alpha}\varphi_{\omega,\vec{k}_\perp}(t,x_{\|},
 \vec{x}_{\perp}).
\label{finite_trans}
\end{equation}
These eigenfunctions satisfy  
\begin{equation}
\varphi_{\omega,\vec{k}_\perp}(-t,-x_{\|},\vec{x}_{\perp}) = 
\varphi_{-\omega,\vec{k}_\perp}^{*}(t,x_{\|},\vec{x}_{\perp}),
\label{cc}
\end{equation}
which is easily verified from the explicit 
forms of these functions \cite{hughes}.  Finally, if $x^{\mu}=(t,x_{\|},\vec{x}_{\perp})$ lies in 
$R$,  $\alpha$ can be analytically continued to the strip $ 0 \leq \Im (\alpha) \leq  +\pi/a$ \cite{fulling}. Then from (\ref{finite_trans}) for $\alpha=-\frac{i\pi}{a}$,
 we get 
\begin{eqnarray}
 \varphi_{-\omega,\vec{k}_\perp}^{*}(t,x_{\|},\vec{x}_{\perp})=e^{-\pi \omega}\varphi_{\omega,\vec{k}_\perp}(t,x_{\|},\vec{x}_{\perp}) \hspace{.5cm}  \mathrm{for} (t,x_{\|},\vec{x}_\perp) \in \mathrm{R}.
\label{finite_trans1}
\end{eqnarray}
 Similarly on the Left Rindler wedge $L$, 
\begin{eqnarray}
 \varphi_{-\omega,\vec{k}_\perp}^{*}(t,x_{\|},\vec{x}_{\perp})=e^{\pi \omega}\varphi_{\omega,\vec{k}_\perp}(t,x_{\|},\vec{x}_{\perp}) \hspace{.5cm}  \mathrm{for} (t,x_{\|},\vec{x}_\perp) \in \mathrm{L}.
\label{finite_trans2}
\end{eqnarray}
The eigenfunctions of  $\mathcal{K}^{R}$ (the boost generator in $R$) are
\begin{equation}
\varphi_{\omega,\vec{k}_\perp, R}(t,x_{\|},\vec{x}_\perp) = i\sqrt{\frac{(2\pi)^{d}(2\omega)}
{|2 \sinh(\pi \omega)|}}[e^{\frac{\pi\omega}{2}}\varphi_{\omega
\vec{k}_\perp}(t,x_{\|},\vec{x}_\perp) - e^{\frac{-\pi\omega}{2}}\varphi_{-\omega,\vec{k}_\perp}^{\ast}(t,x_{\|},\vec{x}_\perp)]. 
\label{eigenkrd}
\end{equation}
For $\omega>0$, these functions are localized in the right Rindler wedge $R$: 
$\varphi_{\omega,\vec{k}_\perp, R}(t,x_{\|},\vec{x}_\perp)=0$ for $(t,x_{\|},\vec{x}_\perp)\in L$. 

Similarly, the eigenfunctions of $\mathcal{K}^{L}$ are
\begin{equation}
\varphi_{\omega,\vec{k}_\perp, L}(t,x_{\|},\vec{x}_\perp) = i\sqrt{\frac{(2\pi)^{d}(2\omega)}
{|2 \sinh(\pi \omega)|}}[e^{\frac{\pi\omega}{2}}\varphi_{-\omega,\vec{k}_\perp}^{\ast}(t,x_{\|},\vec{x}_\perp) - e^{\frac{-\pi\omega}{2}}\varphi_{\omega,\vec{k}_\perp}(t,x_{\|},\vec{x}_\perp)] 
\label{eigenkld}
\end{equation} 
for $\omega>0$ and $\varphi_{\omega,\vec{k}_\perp, L}(t,x_{\|},\vec{x}_\perp)=0$ for $(t,x_{\|},\vec{x}_\perp)\in R$.

These functions $\varphi_{\omega,\vec{k}_\perp, L}(t,x_{\|},\vec{x}_\perp)$ and $\varphi_{\omega,\vec{k}_\perp, R}(t,x_{\|},,\vec{x}_\perp)$ form a complete orthonormal set with 
orthonormality given by \cite{hughes}
\begin{eqnarray}
\left. \begin{array}{l l l}
\langle\varphi_{\omega,\vec{k}_\perp, R}(t,x_{\|},\vec{x}_\perp)|\varphi_{\omega^{\prime},\vec{k}_\perp^{\prime}, L}(t,x_{\|},\vec{x}_\perp)\rangle_{M} = 0   \\
\langle\varphi_{\omega,\vec{k}_\perp, R}(t,x_{\|},\vec{x}_\perp)|\varphi_{\omega^{\prime},\vec{k}_\perp^{\prime}, R}(t,x_{\|},\vec{x}_\perp)\rangle_{M} = (2\pi)^d(2\omega)\delta(\omega-\omega^{\prime})\delta^{d-1}(\vec{k}_\perp-\vec{k}_\perp^{\prime})sgn(\omega) \\
\langle\varphi_{\omega,\vec{k}_\perp, L}(t,x_{\|},\vec{x}_\perp)|\varphi_{\omega^{\prime},\vec{k}_\perp^{\prime}, L}(t,x_{\|},\vec{x}_\perp)\rangle_{M} = (2\pi)^d(2\omega)\delta(\omega-\omega^{\prime})\delta^{d-1}(\vec{k}_\perp-\vec{k}_\perp^{\prime})sgn(-\omega).\\
\end{array}\hspace{0.2cm} \right.
\label{ortho_nor}
\end{eqnarray}
One should not get confused between  $\varphi_{\omega,\vec{k}_\perp}(t,x_{\|},\vec{x}_\perp)$ and $\varphi_{\omega,\vec{k}_\perp, R(L)}(t,x_{\|},\vec{x}_\perp)$ which are completely different 
functions with different properties.

\subsection{Massless Field in $(1+1)$ dimensions}
In $(1+1)$ dimension (d=1), equation (\ref{boost_eve1}) becomes
\begin{eqnarray}
 (x_\| \partial_t + t \partial_{x_\|})\varphi_{\omega}(t,x_{\|})=\omega \varphi_{\omega}(t,x_{\|})\hspace{1cm} -\infty<\omega<\infty
\label{boost_eve}
\end{eqnarray}
and for the massless case the dispersion relation is $(k^{0})^2=(k_\|)^2$. 
For $(1+1)$ dimension, the labels $\vec{k}_\perp $ and  $\vec{x}_\perp $ are absent. We know that 
the sign of energy is an invariant of the restricted Poincar\'e group, and for massless fields the sign of 
$k_{\|}$ (the momentum along $x_\|$)  is invariant under Lorentz boost. Therefore on solving the 
eigenvalue equation (\ref{boost_eve}) we get two positive energy and two negative energy solutions. 
So there are two positive energy solutions :  $\varphi_{\omega,1}(t,x_{\|})$ and 
$\varphi_{\omega,2}(t,x_{\|})$ for positive and negative $x_\|$- momentum respectively whose explicit 
forms are:
\begin{eqnarray}
\varphi_{\omega,1}(t,x_{\|}) &=(2\pi)^{-\frac{1}{2}}\int_{0}^{\infty}\frac{dk_{\|}}{\sqrt{(4\pi k_{\|})}} (k_{\|})^{-i\omega-\frac{1}{2}}e^{-i(k^0 t-k_\| x_\|)} 
\label{k+w}, \\
\varphi_{\omega,2}(t,x_{\|}) &=(2\pi)^{-\frac{1}{2}}\int_{-\infty}^{0}\frac{dk_{\|}}{\sqrt{(4\pi|k_{\|}|)}} |k_{\|}|^{i\omega-\frac{1}{2}}e^{-i(k^0 t-k_\| x_\|)}.
\label{-k+w}
\end{eqnarray}
Similarly there are two negative energy solutions of the equation (\ref{boost_eve}) : $\varphi_{\omega,3}(t,x_{\|})$ and $\varphi_{\omega,4}(t,x_{\|})$
 corresponding to the positive and negative $k_{\|}$ respectively, which are related to the positive energy solutions as
\begin{eqnarray}
 \left. \begin{array}{l l}
 \varphi_{\omega,3}(t,x_{\|})=\varphi_{-\omega,1}^{\ast}(t,x_{\|}), \\
 \varphi_{\omega,4}(t,x_{\|})= \varphi_{-\omega,2}^{\ast}(t,x_{\|}). \\
\end{array}\hspace{0.2cm} \right.
\label{-w}
\end{eqnarray}
The $\varphi_{\omega,\lambda}(t,x_{\|})$ forms a complete orthonormal set of functions : 
$\langle\varphi_{\omega,\lambda}(t,x_{\|})|\varphi_{\omega^{\prime},\lambda^{\prime}}(t,x_{\|})\rangle_{M} = M_{\lambda\lambda^{\prime}}\delta(\omega-\omega^{\prime})$ where $ M_{\lambda\lambda^{\prime}}=\mathrm{diag}(1,1,-1,-1)$.
 
Analogous to (\ref{finite_trans1}) and (\ref{finite_trans2}) we have 
\begin{eqnarray}
 \left. \begin{array}{l l}
  \varphi_{-\omega,\lambda}^{*}(t,x_{\|})=e^{-\pi \omega}\varphi_{\omega,\lambda}(t,x_{\|}) \hspace{.5cm}  \mathrm{for} (t,x_{\|}) \in \mathrm{R}   \\
 \varphi_{-\omega,\lambda}^{*}(t,x_{\|})=e^{\pi \omega}\varphi_{\omega,\lambda}(t,x_{\|}) \hspace{.7cm}  \mathrm{for} (t,x_{\|}) \in \mathrm{L}   \\
\end{array}\hspace{0.2cm} \right\} \quad \text{for} \quad  \lambda=1,2.
\label{finite_trans3}
\end{eqnarray}
Using  (\ref{finite_trans3}), we can now construct $\varphi_{\omega,\lambda, R}(t,x_{\|})$ and 
$\varphi_{\omega,\lambda, L}(t,x_{\|})$, the eigenfunctions of $\hat{\mathcal{K}}^{R}$ and 
$\hat{\mathcal{K}}^{L}$ analogous to (\ref{eigenkrd}) and (\ref{eigenkld}) (we drop the label 
$k_{\perp}$, and include the label $\lambda$) as in \cite{hughes},
\begin{eqnarray}
\varphi_{\omega,\lambda, R}(t,x_{\|})&=&i\sqrt{\frac{(2\pi)(2\omega)}{|2 \sinh(\pi \omega)|}}
\{e^{\frac{\pi\omega}{2}}\varphi_{\omega,\lambda}(t,x_{\|})-e^{\frac{-\pi\omega}{2}}
\varphi_{-\omega,\lambda}^{\ast}(t,x_{\|}\}  \label{global_eigen3} \\
\varphi_{\omega, \lambda,L}(t,x_{\|})&=&i\sqrt{\frac{(2\pi)(2\omega)}{|2\sinh(\pi \omega)|}}
\{e^{\frac{\pi\omega}{2}}\varphi_{-\omega,\lambda}^{\ast}(t,x_{\|})- e^{\frac{-\pi\omega}{2}}
\varphi_{\omega,\lambda}(t,x_{\|})\} \label{global_eigen4} 
\end{eqnarray}
respectively for all $( t,x_\|)$. Again explicit verification using (\ref{finite_trans3}) shows that $\varphi_{\omega,\lambda, R}(t,x_{\|})$ (or $\varphi_{\omega,\lambda, L}(t,x_{\|})$) vanishes in $L$ (or $R$). 

The coordinate representatives of the eigenfunctions of $\hat{\mathcal{K}}$ given by
\begin{eqnarray}
 \varphi_{\omega,1}(t,x_{\|})=(8\pi^2)^{-\frac{1}{2}}\sqrt{\frac{\pi}{\omega \sinh(\pi\omega)}}|
 t-x|^{i\omega}\left\{ 
\begin{array}{l l}
 e^{\pi\omega/2} & (t,x_{\|}) \in R\\
  e^{-\pi\omega/2} & (t,x_{\|}) \in L\\
\end{array} \right.
\label{egn1}
\end{eqnarray}
and
\begin{eqnarray}
 \varphi_{\omega,2}(t,x_{\|})=-(8\pi^2)^{-\frac{1}{2}}\sqrt{\frac{\pi}{\omega \sinh(\pi\omega)}}|t+x|^{-i\omega}\left\{ 
\begin{array}{l l}
 e^{\pi\omega/2} & (t,x_{\|}) \in R\\
  e^{-\pi\omega/2} & (t,x_{\|}) \in L.\\
\end{array} \right.
\label{egn2}
\end{eqnarray}
It can be easily verified that both (\ref{egn1}) and  (\ref{egn2}) satisfy (\ref{cc}). For positive 
$\omega$, the coordinate representatives of the eigenfunctions of $\hat{\mathcal{K}}^{R}$ are
\begin{eqnarray}\varphi_{\omega,1,R}(t,x_{\|})= \left\{
\begin{array}{ll}
 i\hspace{.1cm} \mathrm{sgn} (\omega)\frac{1}{\pi} (t-x_{\|})^{i\omega} & \mathrm{for} (t,x_{\|}) \in R  \\
0 &\mathrm{for} (t,x_{\|}) \in L \\
\end{array} \right.
\end{eqnarray}
and
\begin{eqnarray}\varphi_{\omega,2,R}(t,x_{\|})=\left\{
\begin{array}{ll}
  -i\hspace{.1cm}\mathrm{sgn}(\omega)]\frac{1}{\pi}  (t+x_{\|})^{-i\omega} & \mathrm{for} (t,x_{\|}) \in R  \\
0 & \mathrm{for} (t,x_{\|}) \in L
\end{array} \right.
\end{eqnarray}
The coordinate representatives of the eigenfunctions of $\hat{\mathcal{K}}^{L}$ are
\begin{eqnarray}\varphi_{\omega,1,L}(t,x_{\|})=\left\{
\begin{array}{ll}
 i\hspace{.1cm}\mathrm{sgn}(\omega)\frac{1}{\pi} (t-x_{\|})^{i\omega} & \mathrm{for} (t,x_{\|}) \in L  \\
0 &\mathrm{for} (t,x_{\|}) \in R
\end{array} \right.
\end{eqnarray}
and
\begin{eqnarray}\varphi_{\omega,2,L}(t,x_{\|})=\left\{
\begin{array}{ll}
  -i\hspace{.1cm}\mathrm{sgn}(\omega)\frac{1}{\pi}  (t+x_{\|})^{-i\omega} & \mathrm{for} (t,x_{\|}) \in L \\
0 & \mathrm{for} (t,x_{\|}) \in R.
\end{array} \right.
\end{eqnarray}
The orthonormalization of these complete set of functions are given by
\begin{eqnarray}
&&\langle\varphi_{\omega,\lambda, R}(t,x_{\|},\vec{x}_\perp)|\varphi_{\omega^{\prime},
\lambda^{\prime},L}(t,x_{\|},\vec{x}_\perp)\rangle_{M} = 0,   \nonumber \\
&&\langle\varphi_{\omega,\lambda, R}(t,x_{\|},\vec{x}_\perp)|\varphi_{\omega^{\prime},
\lambda^{\prime} R}(t,x_{\|},\vec{x}_\perp)\rangle_{M} = (2\pi)(2\omega)
\delta_{\lambda\lambda^{\prime}}\delta(\omega-\omega^{\prime})\mathrm{sgn}(\omega), \nonumber \\
&&\langle\varphi_{\omega,\lambda, L}(t,x_{\|},\vec{x}_\perp)|\varphi_{\omega^{\prime},
\lambda^{\prime} L}(t,x_{\|},\vec{x}_\perp)\rangle_{M} = (2\pi)(2\omega)
\delta_{\lambda\lambda^{\prime}}\delta(\omega-\omega^{\prime})\mathrm{sgn}(-\omega).
\end{eqnarray}

\subsection{Massive Field in $(1+1)$ dimensions}
 For massive case in $1+1$ dimensions, (\ref{boost_eve1}) again becomes (\ref{boost_eve}) with the dispersion relation $(k^0)^2-(k_\|)^2=m^2$.
But here the sign of $k_\|$ is not an invariant under Lorentz boost though the sign of energy is 
an invariant of the restricted Poincar\'e group. So we get two solutions of (\ref{boost_eve}), 
one corresponding to positive energy and the other to negative energy.

The positive energy solution is
\begin{equation}
\varphi_{\omega}(t,x_{\|})=(2\pi)^{-\frac{1}{2}}\int_{-\infty}^{\infty}\frac{dk_{\|}}{\sqrt{4\pi} k_{\|}} 
  (k^{0}+k_{\|})^{-i\omega}e^{-i(k^0 t-k_\| x_\|)}
\end{equation}
for $\omega>0$. The coordinate representatives of the positive  energy eigenfunctions of 
$\hat{\mathcal{K}}$ are \cite{gerlach}
\begin{eqnarray}
  \varphi_{\omega}(t,x_{\|})=\left\{
\begin{array}{l l}
 \frac{1}{\sqrt{2}\pi}e^{\pi\omega/2}K_{\frac{i\omega}{a}}(m\sqrt{x_{\|}^{2}-t^{2}})e^{\frac{-i\omega}{a}\tanh(\frac{t}{x_{\|}})} & (t,x_{\|}) \in R, \\
  \frac{1}{\sqrt{2}\pi}e^{-\pi\omega/2}K_{\frac{i\omega}{a}}(m\sqrt{x_{\|}^{2}-t^{2}})e^{\frac{-i\omega}{a}\tanh(\frac{t}{x_{\|}})} & (t,x_{\|}) \in L.\\
\end{array} \right.
\label{eigen7}
\end{eqnarray}
These are known as Minkowski Bessel modes. They form a complete orthonormal set with orthonormalization given by (\ref{orthonormal5}).
It should be noted that the eigenfunctions (\ref{eigen7}) satisfy the relation (\ref{cc}).

As described in Section (\ref{e_b_o}), now we can construct the eigenfunctions of 
$\hat{\mathcal{K}}^{R}$. The eigenfunctions of the generator of Lorentz boost restricted in the $R$ is 
\begin{eqnarray}\varphi_{\omega,R}(t,x_{\|})=\left\{
\begin{array}{ll}
 [\frac{\omega\sinh(\pi\omega/a)}{\pi^{3}a}]^{\frac{1}{2}}K_{\frac{i\omega}{a}}(m\sqrt{x_{\|}^{2}-t^{2}})e^{\frac{-i\omega}{a}\tanh(\frac{t}{x_{\|}})} & \mathrm{for} (t,x_{\|}) \in R, \\
0 &\mathrm{for} (t,x_{\|}) \in L
\end{array}
\right.
\label{eigenfunc}
\end{eqnarray}
for $\omega>0$.  
Similarly we construct $\varphi_{\omega,L}(t,x_{\|})$. These are complete sets of functions which 
satisfy orthonormality condition (\ref{ortho_nor}).

\subsection{Massive Field in $(3+1)$ dimensions}
 In $(3+1)$ dimensions, the eigenvalue equation (\ref{boost_eve1}) is 
\begin{eqnarray}
 \hat{\mathcal{K}}\varphi_{\omega,\vec{k}_\perp}(t,x_{\|},\vec{x}_{\perp})=\omega
 \varphi_{\omega,\vec{k}_\perp}(t,x_{\|},\vec{x}_{\perp})   \hspace{1cm} -\infty<\omega<\infty
\label{boost_eve2} 
\end{eqnarray}
where  $\omega$ is a continuous real eigenvalue and $\vec{k}_\perp$ is the two-dimensional 
vector component of spatial momentum $(\vec{k}=(k_\|,\vec{k}_\perp))$.

The positive energy solution is
\begin{equation}
\varphi_{\omega,\vec{k}_{\perp}}(t,x_{\|},\vec{x}_\perp)=(2\pi)^{-\frac{1}{2}}\int_{-\infty}^{\infty}
\frac{dk_{\|}}{\sqrt{4\pi} k_{\|}} (k^{0} +  k_{\|})^{-i\omega} e^{-i(k^0 t-k_{\|} x_{\|}+\vec{k}_{\perp}
\cdot\vec{x}_{\perp})}
\end{equation}
for $\omega>0$. The coordinate representatives can be obtained as in \cite{gerlach}
\begin{eqnarray}
  \varphi_{\omega,\vec{k}_{\perp}}(t,x_{\|},\vec{x}_\perp)=\left\{
\begin{array}{l l}
\frac{1}{\sqrt{2}\pi}e^{\pi\omega/2}K_{\frac{i\omega}{a}}(m\sqrt{x_{\|}^{2}-t^{2}})e^{\frac{-i\omega}{a}\tanh(\frac{t}{x_{\|}})+i\vec{k}_{\perp}\cdot\vec{x}_{\perp}} & {\rm for} \,\,(t,x_{\|},\vec{x}_\perp) \in R,\\
\frac{1}{\sqrt{2}\pi}e^{-\pi\omega/2}K_{\frac{i\omega}{a}}(m\sqrt{x_{\|}^{2}-t^{2}})e^{\frac{-i\omega}{a}\tanh(\frac{t}{x_{\|}})+i\vec{k}_{\perp}\cdot\vec{x}_{\perp}} & {\rm for} \,\,(t,x_{\|},\vec{x}_\perp) \in L.
 \end{array} \right.
\label{eigen8}
\end{eqnarray}
These form a complete orthonormal set with orthonormalization given by (\ref{orthonormal5}). 
The functions (\ref{eigen8}) obey (\ref{cc}). 

The eigenfunctions of $\hat{\mathcal{K}}^{R}$ (the generator of Lorentz boost restricted in the $R$) 
are 
\begin{eqnarray}\varphi_{\omega,\vec{k}_{\perp},R}(t,x_{\|},\vec{x}_\perp)=\left\{
\begin{array}{ll}
 [\frac{\omega\sinh(\pi\omega/a)}{a}]^{\frac{1}{2}}K_{\frac{i\omega}{a}}(m\sqrt{x_{\|}^{2}-t^{2}})e^{\frac{-i\omega}{a}\tanh(\frac{t}{x_{\|}})+i\vec{k}_{\perp}\cdot\vec{x}_{\perp}} & \mathrm{for} (t,x_{\|},\vec{x}_\perp) \in R,  \\
0 &\mathrm{for} (t,x_{\|},\vec{x}_\perp) \in L
\end{array}
\right.
\label{eigenfunc2}
\end{eqnarray}
for $\omega>0$ and where $\kappa=\sqrt{m^2+\vec{k}_\perp\cdot\vec{k}_\perp}$. Similarly we 
construct $\varphi_{\omega,\vec{k}_{\perp},L}(t,x_{\|},\vec{x}_\perp)$, the eigenfunctions of 
$\hat{\mathcal{K}}^{L}$. These functions satisfy orthonormality condition (\ref{ortho_nor}).

\end{document}